\documentstyle[12pt]{article}

\begin{document}

\setcounter{page}{1}

\pagestyle{plain}
\vspace{1cm}
\begin{center}
\Large{\bf Black body radiation in a model universe with large extra dimensions and quantum gravity effects}\\
\small \vspace{1cm}
{\bf Kourosh Nozari$^{a,b,}$\footnote{knozari@umz.ac.ir}},\quad{\bf S. F. Anvari$^{a,}$\footnote{f.anvari@stu.umz.ac.ir}}\quad and \quad{\bf A. S. Sefiedgar$^{a,}$\footnote{a.sefiedgar@umz.ac.ir}} \\
\vspace{0.5cm} {$^{a}$Department of Physics, Faculty of Basic Sciences,
University of Mazandaran,\\
P. O. Box 47416-95447, Babolsar, IRAN}\\
\vspace{0.35cm}
{$^{b}$ Research Institute for Astronomy and Astrophysics of Maragha
(RIAAM)\\ P. O. Box 55134-441, Maragha, Iran}
\end{center}
\vspace{1.5cm}
\begin{abstract}
We analyze the problem of black body radiation in the presence of
quantum gravity effects encoded in modified dispersion relations and
in a model universe with large extra dimensions. In this context,
modified form of Planck distribution, Jeans number, equipartition
theorem, spectral energy density, Stefan-Boltzmann law and Wien's
law are found and the corresponding results are interpreted. Then,
entropy and specific heat of black body radiation are obtained in
this setup. Finally, the modified form of Debye law and Dulong-Petit
law are investigated in this framework.
\end{abstract}
\newpage

\section{Introduction}
String theory as a possible candidate for quantum gravity proposal
implies existence of extra spacetime dimensions. In fact, existence
of possible extra spatial dimensions has been a principal key to
address the long-lived Hierarchy problem [1,2]. The first attempt to
adopt extra dimensions in theoretical physics was made by Kaluza and
Klein to unify gravity with electromagnetism in a 5-dimensional
spacetime model. Currently the idea of having extra dimensions has
been widely accepted by theoretical physics community and there is
also hope to detect some footprints of these extra dimensions in
experiments such as the LHC. After revolutionary report on current
positively accelerating phase of the universe expansion [3],
possible modification of General Relativity has been in the center
of some important research programs. One may modify the geometric
part of the Einstein's equations by allowing to have extra spatial
dimensions in the framework of braneworld scenarios [4]. These
attempts have been applied successfully to explain the issue of dark
energy and interpretation of new astrophysical data. These are just
parts of possible motivation to introduce extra dimensional
scenarios in theoretical physics. Nevertheless, the presence of
extra dimensions would remain concealed some how, may be because the
extra dimensions are compactified to small radius with the size
around Planck scale. Extra dimensional scenarios open also new
perspectives, at least phenomenologically, to quantum gravity
proposal. A common feature of most approaches to quantum gravity,
such as string theory, noncommutative geometry, loop quantum gravity
and Doubly Special Relativity, is modification of the standard
dispersion relation (see for instance [5,6,7] and references
therein). In fact, in the study of loop quantum gravity, Doubly
Special Relativity and models based on non-commutative geometry
there has been strong indication to modify the standard
energy-momentum dispersion relation. In this framework, issues such
as possible violation of the Lorentz invariance in quantum gravity
regime are studied too (see [8] and references therein). Recently
modified dispersion relations are formulated in model universes with
extra dimensions by focusing on black hole thermodynamics in these
models [9]. Here we focus on another important issue: the block body
radiation in a model universe with large extra dimensions and by
incorporating possible quantum gravity effects encoded in modified
dispersion relations. A black body is an ideal body which allows the
whole of the incident radiation to pass into itself and absorbs
within itself this whole incident radiation. One cannot ignore the
significance of study of black body spectrum and its possible
modification since it could be as a reliable source to give some
information about the remnant of early universe radiation which
could be detectable today as cosmic background microwaves (CMB). In
fact, the cosmic microwave background radiation observed today is
the most perfect black body radiation ever observed in nature, with
a temperature of about $2.725$ K. It is a snapshot of the radiation
at the time of decoupling between matter and radiation in the early
stages of the universe evolution. Therefore, any modification of
black body spectrum due to quantum gravity effects can open new
windows to know more about quantum gravitational features of the
early universe through study of CMB spectrum. This sort of
investigation has been considered by some authors recently [10].
Also formulation of black body radiation in model universe with
extra dimensions, but in the absence of quantum gravity effects, has
been considered by some authors [11]. The crucial feature of our
work is the combination of MDR (as a phenomenological outcome of
quantum gravity proposal) with possible existence of extra
dimensions in problem of black body radiation. This study is an
attempt to take a small step toward a deeper understanding, at least
phenomenologically, of quantum gravity in a model universe with
large extra dimensions. In this framework we investigate the impact
of modified dispersion relations (MDR) on black body spectrum in a
model universe with large extra dimensions. We establish modified
form of the Planck distribution, equipartition theorem,
Stefan-Boltzmann law and Wien's law in this setup. Also modified
Debye law and modified Dulong-Petit law are obtained and the
corresponding results are interpreted. Then we study entropy and
specific heat of black body radiation in this framework. The
important ingredient of these modified quantities is that these
modifications are themselves temperature-dependent. So, it seems
that quantum gravity effects encoded in modified dispersion
relations are temperature-dependent for black body radiation and
also related issues. This feature may be related to the foamy
structure of space-time at the quantum gravity level, and taking the
quantum structure of spacetime to be as a granular media which has
some interaction with the events which are happened in it. So, these
effects can be attributed to some yet not well-understood features
of quantum spacetime at the Planck scale. We note also that the
assumption of black body in a solid box, to some extent, helps us to
know more about the modified terms within MDR in vibrating solids.
Therefore the modified heat capacity of solids would be achieved
through Dulong-petit model in high temperature [11].

In which follows, we set $\hbar =k_B=c=1$ for simplicity.

\section{Black body radiation in a model universe with extra dimensions}

In order to study a black body radiation and it's features in a
D-dimensional space (D spatial dimensions), we consider a conducting
cavity with the shape of a D-dimensional cube of side $L$. This
cavity is filled with electromagnetic radiation which is in thermal
equilibrium with walls at temperature $T$ and is linked with the
outside by a small hole. The electromagnetic radiation inside the
cavity can be treated as standing waves. We are going to seek for
the functional form of energy density for one-dimensional cavity
that is placed in a D-dimensional universe (see [11] for a
discussion about validity of this assumption). The most proper
method for investigating the black body radiation is deriving the
total number of modes in order to obtain the energy density. Let us
write the components of the electric field in a cavity by choosing a
system of orthogonal coordinates with origin at one of the cavity's
vertices. To satisfy boundary conditions, at the walls of the cavity
the parallel components of the electric field must vanish. Taking
these considerations into account, we have the following relation in
terms of wave number
\begin{equation}
n_i=\frac{k_i}{\pi}L,
\end{equation}
where $k_i$ is the component of wave vector and $n_i=1,2,...$
represents the possible modes of vibration. The total number of
modes in one-dimensional space is given by
\begin{equation}
N(k_i) dk_i=\frac{L}{\pi} dk_i,
\end{equation}
which can be converted to the following form in a spacetime with $D$
spatial dimensions
\begin{equation}
N(k)dk=(D-1)\frac{V} {2^{D} \pi^{D}} dV_k,
\end{equation}
where $V=L^D$ is the volume of the cavity and $dV_k$ is the
infinitesimal element of volume in the $k$-space [11]. Coefficient
$(D-1)$ is related to polarization states and the presence of $2^D$
in the denominator of equation (3) is due to the fact that we should
consider only the positive part of the volume in $k$-space. The
D-dimensional element of volume in terms of the frequency $\nu$ is given by
\begin{equation}
dV_\nu=2\frac{\pi^\frac{D}{2}}{\Gamma(\frac{D}{2})} ({2\pi})^D\,
\nu^{(D-1)} d\nu.
\end{equation}
By using Eq. (4), one can rewrite Eq. (3) in $\nu$-space as
follows
\begin{equation}
N(\nu) d\nu=2 (D-1) V \frac{\pi^{\frac{D}{2}}}{ \Gamma(\frac{D}{2})}
\nu^{(D-1)} d\nu\,.
\end{equation}
This relation expresses the total number of modes with frequencies
between $\nu$ and $\nu+d\nu$.

In the next section, by using the average energy per mode we will
construct the spectral energy density to explain black body
radiation in the presence of quantum gravity effects in a spacetime
with $D$ spatial dimensions.

\section{MDR-modified black body radiation in a model universe with $D$ spatial dimensions}

The modified dispersion relation in a model universe with $D$
spatial dimensions can be written in the following form (see [5,9]
for more details)
\begin{equation}
{\vec p}^2=f(E,m;L_P)\simeq E^2 - \mu^2  + \alpha {L}_p^2 E^4
+\alpha' {L}_p^4 E^6 + O({L}_p^6 E^8)\,,
\end{equation}
where $L_P$, the Planck length in spacetime with $D$ spatial
dimensions, depends on spacetime dimensionality (see [9] for
instance), and $f$ is a function that gives the exact dispersion
relation [5]. On the right hand side of equation (6) we have adopted
a Taylor-series expansion for $E\ll \frac{1}{L_{P}}$. The parameter
$\alpha_i$ is dimensionless and takes different values depending on
the details of quantum gravity candidates. The parameter $m$ is the rest energy of the particle and it is different from the mass parameter $\mu$. we
expect $\mu \neq m$ if the coefficients
$\alpha_i$ do not all vanish [5]. \\
Based on the analysis of the previous section, the number of modes
in the momentum interval between $p$ and $p+dp$, is given by
\begin{equation}
N(p)dp=(D-1)V
\frac{2\pi^\frac{D}{2}}{\Gamma(\frac{D}{2})}p^{D-1}dp\,.
\end{equation}
To find the modified form of the total number of electromagnetic
modes in MDR framework, we start with differentiating and applying a
Taylor expansion to Eq. (6) to obtain
\begin{equation}
 dp=dE \bigg[ 1+\frac{3}{2}\alpha L_p^2 E^2 +\Big(-\frac{5}{8}\alpha^2+\frac{5}{2}\alpha'\Big)L_p^4
 E^4\bigg]\,,
\end{equation}
where we have saved only terms up to the forth order of the Planck
length and the rest mass has been neglected. Substituting the
expressions for $p$ and $dp$, we find
$$N(E)dE=(D-1) V \frac{2\pi^\frac{D}{2}}{\Gamma(\frac{D}{2})}{E}^{D-1}
\bigg(1+\alpha L_{p}^2 E^2+\alpha' L_{p}^4
E^4\bigg)^{\frac{({D-1})}{2}}\times$$
\begin{equation}
\bigg[1+\frac{3}{2} \alpha L_{p}^2 E^2+\Big(-\frac{5}{8} \alpha^2
+\frac{5}{2}\alpha'\Big)L_{p}^4 E^4\bigg]dE.
\end{equation}
This is the modified number of electromagnetic modes in a cavity and
in a model universe with $D$ spatial dimensions. Note that the
average energy per mode is independent on the dimensionality of
space and it can be given by (see for instance [11])
\begin{equation}
\bar{E}=\frac{\nu}{e^{\frac{\nu}{T}}-1}\,.
\end{equation}
Then, the modified spectral energy density is\\
$$\rho(\nu)d\nu=(D-1)\frac{2\pi^\frac{D}{2}}{\Gamma(\frac{D}{2})}\nu^D
\bigg(\frac{1}{e^{\frac{\nu}{T}}-1}\bigg)
\bigg\{1+\Big(\frac{3}{2}+\frac{(D-1)}{2}\Big)\alpha L_{p}^2
\nu^2+$$
\begin{equation}
\bigg[\alpha^2\Big(\frac{-5}{8}+\frac{3}{2}\frac{(D-1)}{2}+\frac{(D-1)(D-3)}{8}\Big)
+\alpha'\Big(\frac{5}{2}+\frac{(D-1)}{2}\Big)\bigg]L_{p}^4
\nu^4\bigg\}d\nu.
\end{equation}\\
This is the modified Planck distribution. The influence of modified
dispersion relation on spectral energy density in a model universe
with large extra dimensions is obviously considerable. Figure $1$
shows the variation of the spectral energy density versus $\nu$ in
Planck temperature for a model universe with $D$ spatial dimensions
but without MDR effects. This figure shows that when the number of
spatial dimensions increases, the frequency at which the
distribution maximizes (the location of the distribution peak) will
attain a noticeable shift toward higher frequencies. Also the
distribution becomes wider and its height \emph{decreases} by
increasing $D$. This shows that by increasing $D$, the number of
modes located around the most probable frequency increases
considerably. Figure $2$ depicts the variation of the
\emph{modified} spectral energy density versus $\nu$ in $D$-spatial
dimensions when quantum gravity effects are taken into account
through MDR and temperature is fixed at the Planck temperature.
While the location of peaks in this case are $D$-dependent in the
same way as in Fig. $1$, the height of picks \emph{increases} by
increasing $D$ and the distribution width increases also with $D$.
So, there is a sharp difference between the results obtained in the
presence of quantum gravity effects and the ones obtained in the absence of quantum gravity effects. This feature
may provide in principle a direct clue to test the situation in the
lab.

\begin{figure}[htp]
\begin{center}
\includegraphics{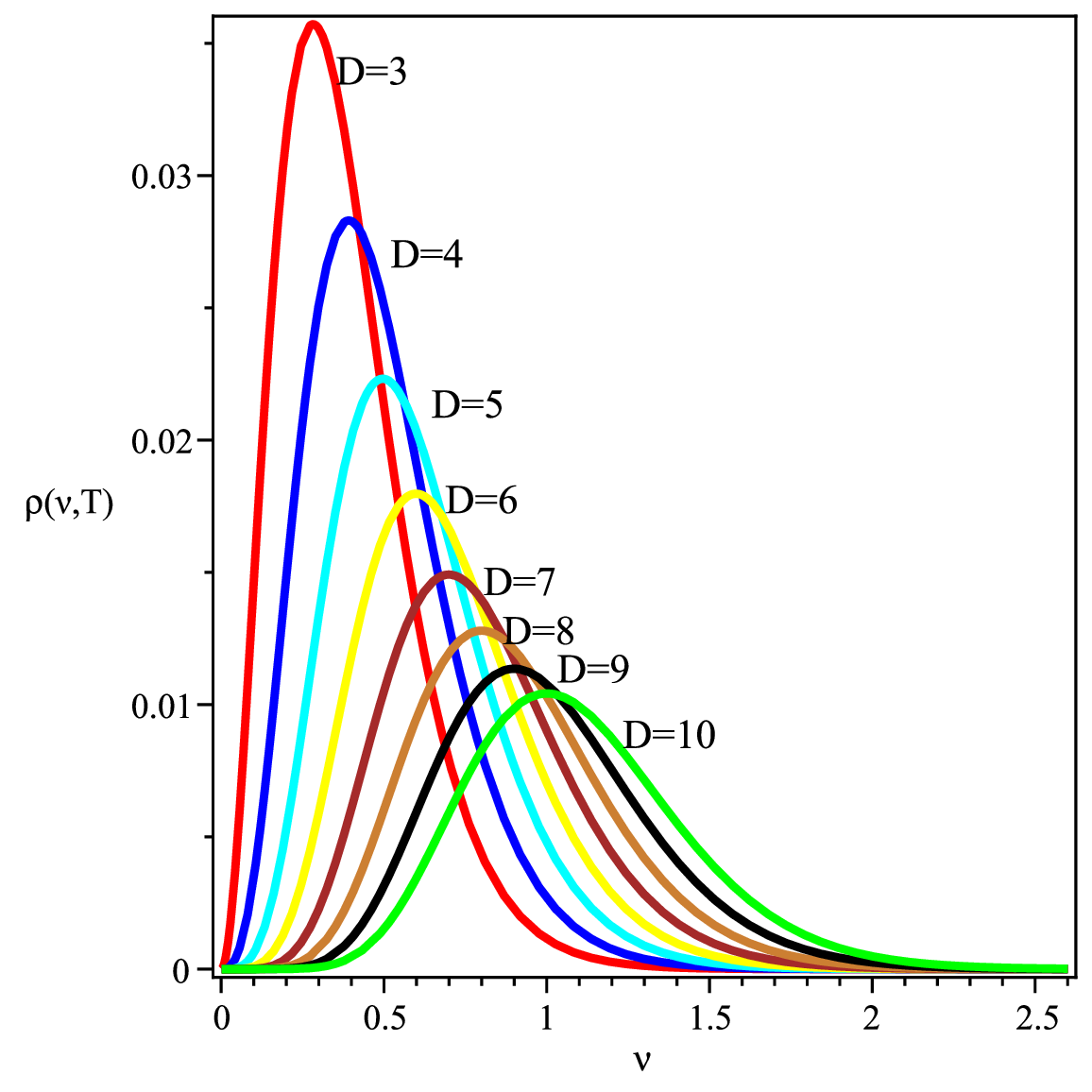}
\end{center}
\vspace{8cm} \caption{\scriptsize{Spectral energy density as a
function of frequency in a model universe with $D$ spatial
dimensions and without MDR effects. }}
\end{figure}

\begin{figure}[htp]
\begin{center}
\includegraphics{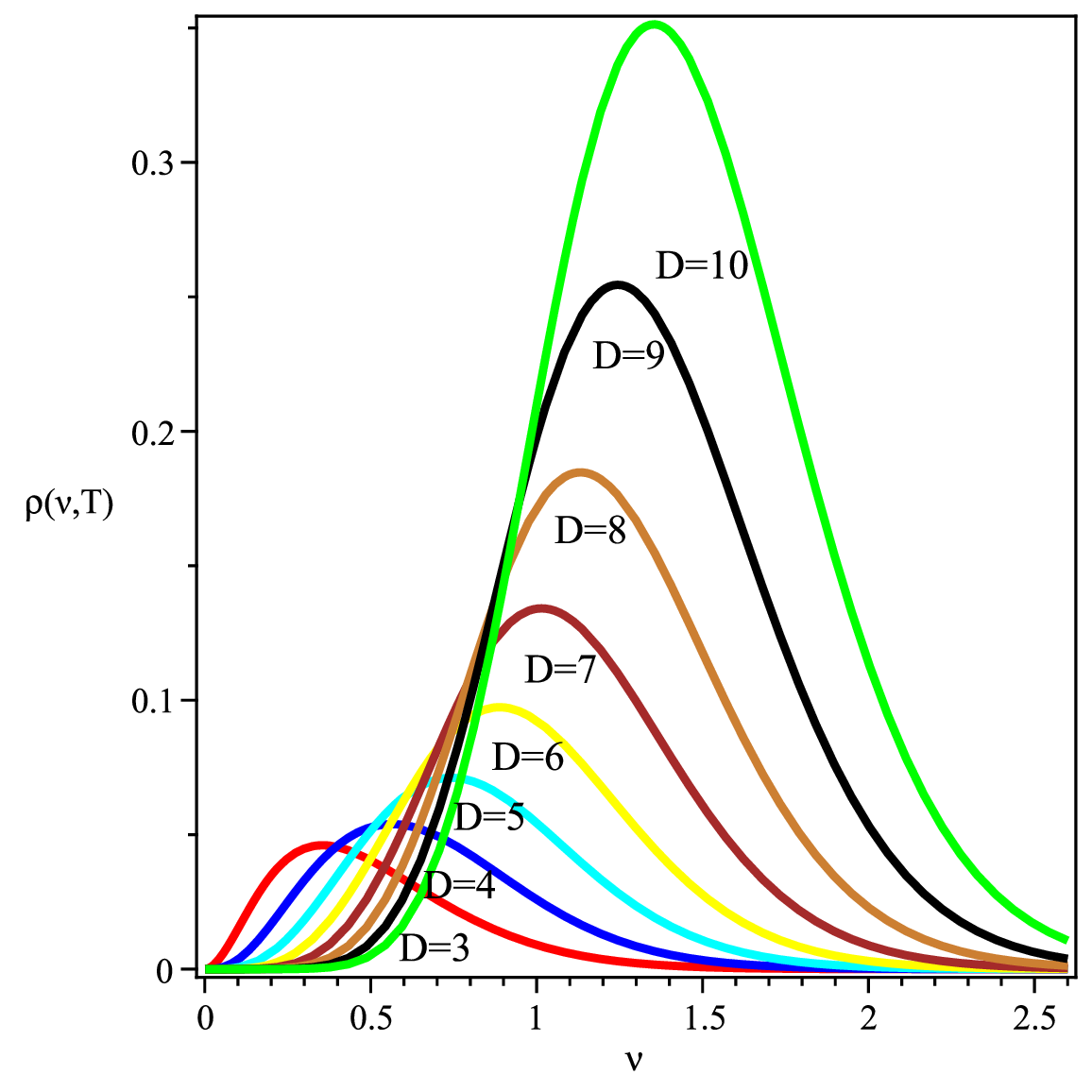}
\end{center}
\vspace{7cm} \caption{\scriptsize{Spectral energy density as a
function of frequency in a model universe with $D$ spatial
dimensions and in the presence of quantum gravity effects encoded in
MDR. }}
\end{figure}

Comparison between these two figures shows that when quantum
gravitational effects are taken into account, the frequency at which
the distribution maximizes will attain a noticeable shift toward
high frequencies. That is because of the impact of quantum
gravitational terms on spectral energy density. Now we explain an
important achievement of our analysis through figures 3 and 4. These
two figures show the spectral energy density as a function of
frequency with and without incorporation of quantum gravity effects
for $D=3$ and $8$ respectively. By incorporation of quantum gravity
effects via MDR, in each figure the frequency at which the
distribution maximizes has significant shift toward the higher
frequencies and the height of the distribution increases too.
Comparison among these two figures shows that as the number of
spatial dimensions increases, the difference between two peaks
(frequency shift) increases the same as the difference between the heights of peaks. Based on these figures, the effects of quantum gravity
are enhanced as the number of spatial dimension's of the universe
increases.
\begin{figure}[htp]
\begin{center}
\includegraphics{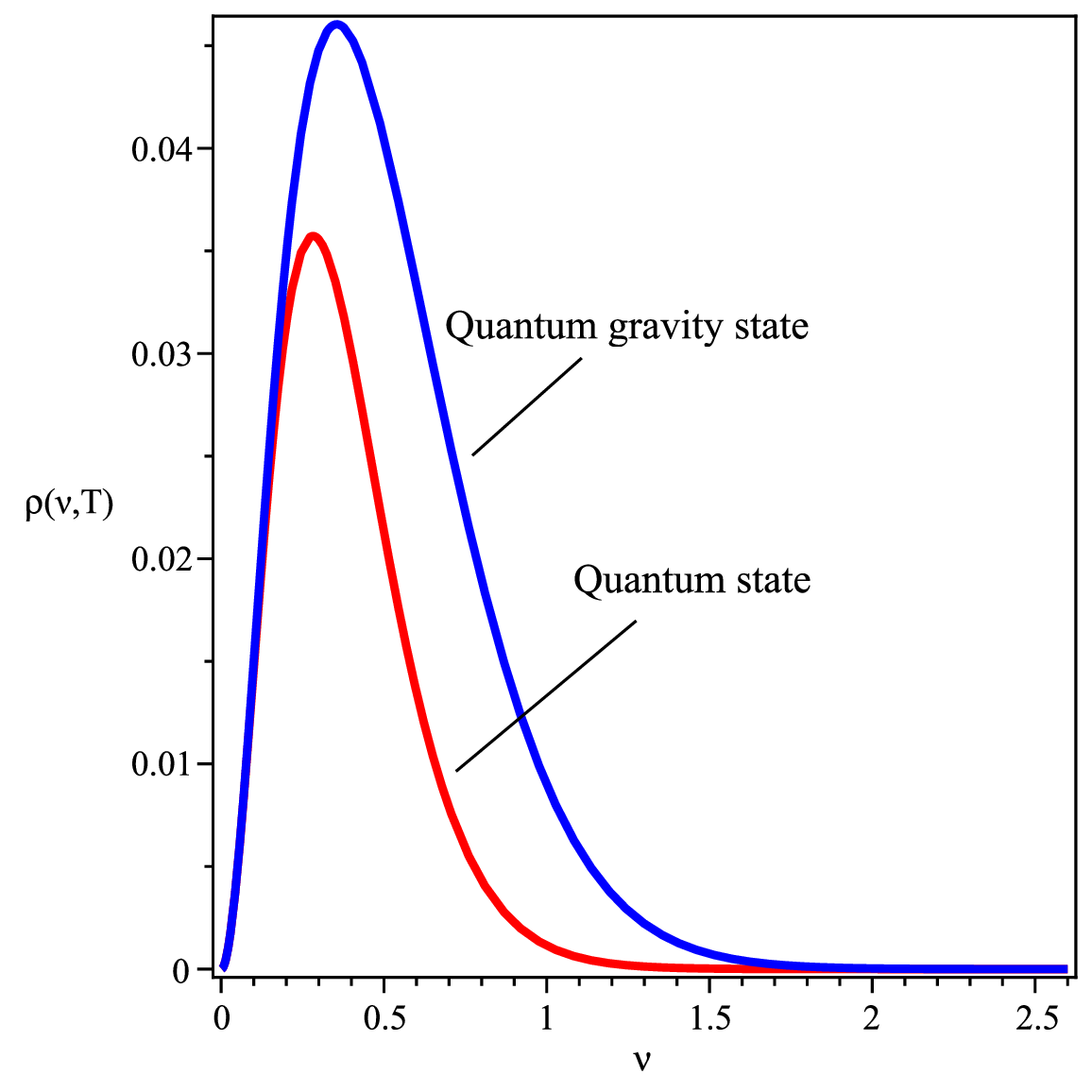}
\end{center}
\vspace{5cm} \caption{\scriptsize{Spectral energy density as a
function of frequency for D=3 in two different regimes. }}
\end{figure}

\begin{figure}[htp]
\begin{center}
\includegraphics{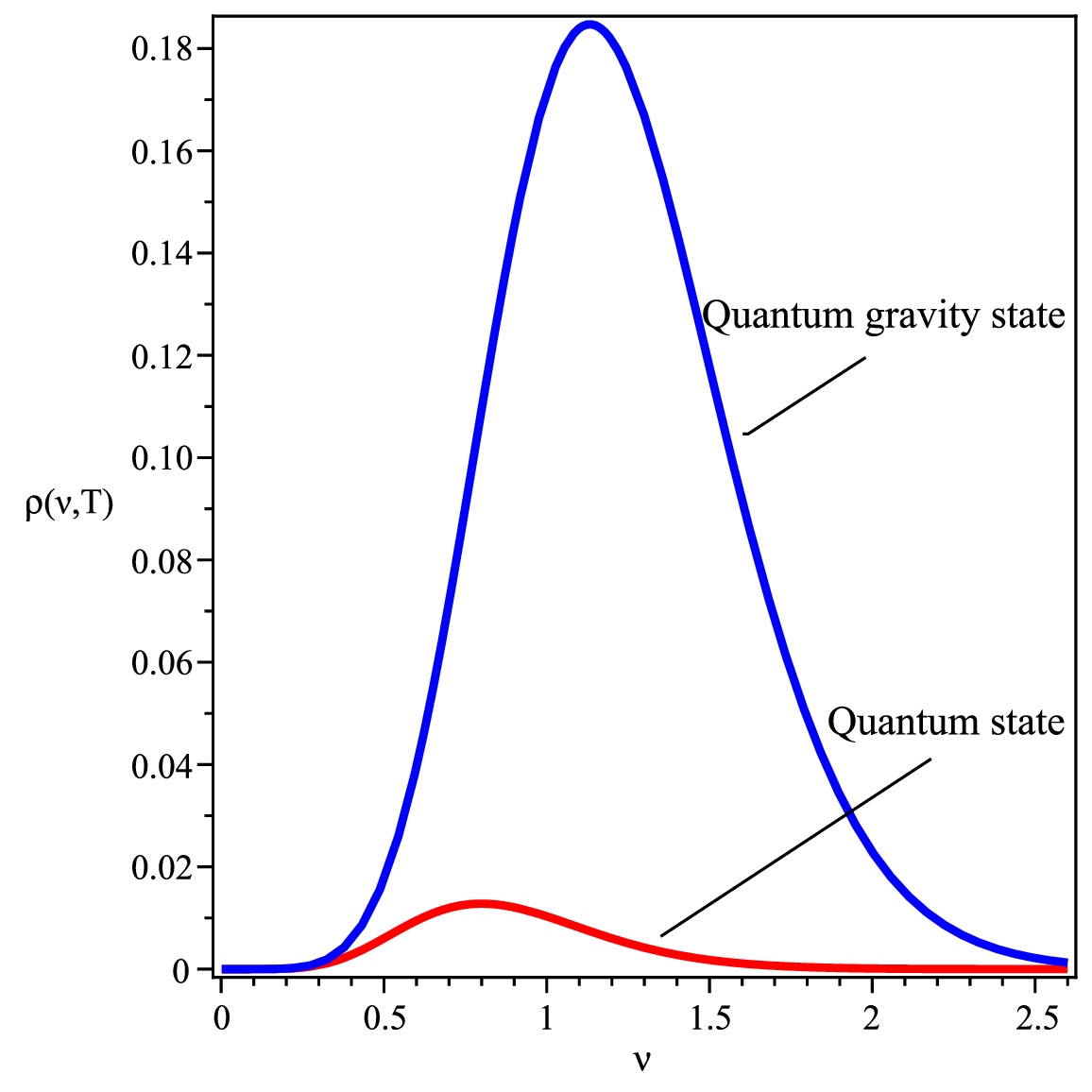}
\end{center}
\vspace{6.5cm} \caption{\scriptsize{Spectral energy density as a
function of frequency for D=8 in two different regimes.}}
\end{figure}

\newpage
In 4-dimensional space-time, classical equipartition theorem states
that average energy per electromagnetic modes of black-body
radiation is $T$ (with $k_{B}=1$). In ordinary quantum mechanics,
this average is given by\, $\frac{\nu}{e^{\frac{\nu}{T}}-1} $.\, Now
in a model universe with large extra dimensions and in the presence
of quantum gravitational effects encoded in MDR, this average
generalizes to
$$\bar{E}=\bigg(\frac{\nu}{e^{\frac{\nu}{T}}-1}\bigg)\bigg[1+\bigg(\frac{3}{2}+\frac{(D-1)}{2}\bigg)\alpha
L_p^2
\nu^2+$$
\begin{equation}
\bigg(\alpha^2(\frac{-5}{8}+\frac{3}{2}\frac{(D-1)}{2}+\frac{(D-1)(D-3)}{8})
+\alpha'(\frac{5}{2}+\frac{(D-1)}{2})\bigg)L_{p}^4 \nu ^4\bigg].
\end{equation}
This is a generalization of the equipartition theorem to quantum
gravitational regime in a model universe with large extra dimensions.\\

\subsection{Modified Rayleigh-Jeans law}
The classical Rayleigh-Jeans law describes the spectral energy
density of electromagnetic waves in black body radiation, which fits
with experimental results at low frequencies, while in high
frequencies it tends to infinity known historically as the
Ultraviolet Catastrophe. The modified form of the Rayleigh-Jeans law
in our framework is written as \\
$$\rho(\nu,T)=\Bigg((D-1)\frac{2\pi^\frac{D}{2}}{\Gamma(\frac{D}{2})}\nu^{D-1}
\Bigg)\Bigg\{1+\Big(\frac{3}{2}+\frac{(D-1)}{2}\Big)\alpha L_{p}^2
\nu^2+$$
\begin{equation}
\bigg[\alpha^2\Big(\frac{-5}{8}+\frac{3}{2}\frac{(D-1)}{2}+\frac{(D-1)(D-3)}{8}\Big)
+\alpha'\Big(\frac{5}{2}+\frac{(D-1)}{2}\Big)\bigg]L_{p}^4
\nu^4\Bigg\}\,T \,.
\end{equation}
Thus, the modified Jean's number can be written as \\
$${\cal{N}}_{J}=\Bigg((D-1)\frac{2\pi^\frac{D}{2}}{\Gamma(\frac{D}{2})}\nu^{D-1}
\Bigg)\Bigg\{1+\Big(\frac{3}{2}+\frac{(D-1)}{2}\Big)\alpha L_{p}^2
\nu^2+$$
\begin{equation}
\bigg[\alpha^2\Big(\frac{-5}{8}+\frac{3}{2}\frac{(D-1)}{2}+\frac{(D-1)(D-3)}{8}\Big)
+\alpha'\Big(\frac{5}{2}+\frac{(D-1)}{2}\Big)\bigg]L_{p}^4
\nu^4\Bigg\}\,.
\end{equation}
Assuming $\alpha_{i}$ are positive quantities, the Jeans's number as
number density of modes in a given frequency increases in the
presence of quantum gravity effects. We note that since the
Reyleigh-Jeans theory has a classical basis, we have used $k_{B}T$
(with $k_{B}=1$) as mean energy per mode in the above calculations.
Nevertheless, in a general framework we should take into account the
modification of equipartition theorem and use equation (11) as our
starting point. Fortunately, the result is the same as equation (14).

\subsection{Modified Stefan-Boltzmann law}
In ordinary quantum mechanics with $D=3$, the total energy density
of black body radiation is proportional to the forth power of
temperature, that is, $\rho(T)=aT^4$. \\ In a higher dimensional
world, when ordinary quantum mechanics is taken into account, the
total energy density is $\rho(T)=a_D T^{D+1}$ [11]. Now, in a model
universe with large extra dimensions and in the presence of quantum
gravity effects encoded in the MDR, we integrate equation (11) over
frequency to find

$$\rho(T)=\frac{(D-1)2{\pi}^\frac{D}{2}
T^{D+1}}{\Gamma(\frac{D}{2})}\Bigg\{\zeta(D+1)
\Gamma(D+1)+\bigg[\frac{3}{2}+\frac{(D-1)}{2}\bigg]\alpha {L_p}^2
T^2 \zeta(D+3) \Gamma(D+3)$$
\begin{equation}
+\bigg[\alpha^2\Big(-\frac{5}{8}+\frac{3}{2}\frac{(D-1)}{2}+\frac{(D-1)(D-3)}{8}\Big)
+\alpha'\Big(\frac{5}{2}+\frac{D-1}{2}\Big)\bigg]{L_p}^4 T^4
\zeta(D+5)\Gamma(D+5)\Bigg\}\,.
\end{equation}
In comparison with $\rho(T)=a_D T^{D+1}$, the coefficient $a_D$ in
the presence of quantum gravity effects is given by

$$a_D=\frac{(D-1)2{\pi}^\frac{D}{2}}{\Gamma(\frac{D}{2})}\Bigg\{\zeta(D+1)
\Gamma(D+1)+\bigg(\frac{3}{2}+\frac{(D-1)}{2}\bigg)\alpha {L_p}^2
T^2 \zeta(D+3) \Gamma(D+3)$$
\begin{equation}
+\bigg[\alpha^2\Big(-\frac{5}{8}+\frac{3}{2}\frac{(D-1)}{2}+\frac{(D-1)(D-3)}{8}\Big)
+\alpha'\Big(\frac{5}{2}+\frac{D-1}{2}\Big)\bigg]{L_p}^4 T^4
\zeta(D+5)\Gamma(D+5)\Bigg\},
\end{equation}
where $\zeta$ and $\Gamma$ are the Riemann Zeta function and gamma
function respectively. The \emph{Radiancy} is defined by the total
energy emitted by a black body per unit area per unit time. The
energy density $\rho(T)$ is proportional to the Radiancy $R(T)$ by a
pure geometric factor [11]
\begin{equation}
R(T)=\frac{\Gamma(\frac{D}{2})}{\Gamma\frac {(D+1)}{2}}
\frac{1}{2\sqrt{\pi} } \rho(T)\,.
\end{equation}
Therefore, the modified Stefan-Boltzmann law may be expressed as
 $$R(T)=\pi^{\frac{D-1}{2}} \frac{(D-1)}{\Gamma(\frac{D+1}{2})}
T^{D+1}\Bigg\{\zeta(D+1)
\Gamma(D+1)+\bigg(\frac{3}{2}+\frac{(D-1)}{2}\bigg)\alpha {L_p}^2
T^2 \zeta(D+3) \Gamma(D+3)$$
\begin{equation}
+\bigg[\alpha^2\Big(-\frac{5}{8}+\frac{3}{2}\frac{(D-1)}{2}+\frac{(D-1)(D-3)}{8}\Big)
+\alpha'\Big(\frac{5}{2}+\frac{D-1}{2}\Big)\bigg]{L_p}^4 T^4
\zeta(D+5)\Gamma(D+5)\Bigg\}\,.
\end{equation}
Now, $\sigma_M$ as the modified Stefan-Boltzmann's factor in a model
universe with $D$ spatial dimensions and in the presence of MDR effects can be written as \\
$$\sigma_M=\pi^{\frac{D-1}{2}} \frac{(D-1)}{\Gamma(\frac{D+1}{2})}
\Bigg\{\zeta(D+1)
\Gamma(D+1)+\bigg(\frac{3}{2}+\frac{(D-1)}{2}\bigg)\alpha {L_p}^2
T^2 \zeta(D+3) \Gamma(D+3)$$
\begin{equation}
+\bigg[\alpha^2\Big(-\frac{5}{8}+\frac{3}{2}\frac{(D-1)}{2}+\frac{(D-1)(D-3)}{8}\Big)
+\alpha'\Big(\frac{5}{2}+\frac{D-1}{2}\Big)\bigg]{L_p}^4 T^4
\zeta(D+5)\Gamma(D+5)\Bigg\}.
\end{equation}
We note that the modified Stefan-Boltzmann's factor in this
framework is not constant and varies with temperature. This
temperature dependence of modifications is a generic feature and
probably has its origin in the quantum fluctuation of background
metric in quantum gravity regime.

\subsection{Modified Wien's law}
In ordinary quantum mechanics and in three space dimensions, there
is an inverse relationship between the wavelength of the peak of the
emission distribution of a black body and it's temperature, which is
called Wien's displacement law. This law is expressed as
$\lambda_{max}=\frac{b}{T}$, where $b$ is Wien's constant and $T$ is
the absolute temperature of the black body when $\lambda_{max}$ is
the distribution peak wavelength. In a higher dimensional universe
and by incorporating the quantum gravity effects, we should rewrite
the modified energy density, Eq. (11), as a function of wavelength.
Then we find
$$\rho_\lambda(T)=(D-1)\frac{2\pi^\frac{D}{2}}{\Gamma(\frac{D}{2})}
\frac{1}{\lambda^{(D+2)} (e^{\frac{1}{\lambda T}}-1)}
\Bigg\{1+\Big(\frac{3}{2}+\frac{(D-1)}{2}\Big)\alpha L_{p}^2
\frac{1}{\lambda^2}+$$
\begin{equation}
\bigg[\alpha^2\Big(\frac{-5}{8}+\frac{3}{2}\frac{(D-1)}{2}+\frac{(D-1)(D-3)}{8}\Big)+\alpha'\Big(\frac{5}{2}+\frac{(D-1)}{2}\Big)\bigg]L_{p}^{4}
\frac{1} {\lambda^4}\Bigg\}\,.
\end{equation}
In order to find the wavelength of maximum emission, we
differentiate $\rho_{\lambda}(T)$ with respect to $ \lambda$ (up to
the second order of $L_{P}$), and then by using the approximation
$e^{\frac{1}{\lambda T}}\approx1+\frac{1}{\lambda T}$ and putting
the derivative equal to zero, we find
$$ \lambda_{max}=-\frac{1}{3A}+\frac{\Big(36 BA-108 F A^2-8+12 \sqrt{3}A \sqrt{4 B^{3} A-B^{2} -18 BA
F+27 F^{2} A^{2}+4 F }\Big)^{\frac{1}{3}}}{6 A}$$
\begin{equation}
-\frac{2}{3} \frac {3 B A-1} {A \Big(36 BA-108 F A^2-8+ 12 \sqrt{3}A
\sqrt{4 B^3 A-B^2 -18 B A F+27 F^2 A^2+4 F}\Big)^{\frac{1}{3}}},
\end{equation}
where $A$, $B$ and $F$ are defined as

$$A=-T(D+1),\,\, B=-T(D+3)(\frac {3}{2}+\frac{(D-1)}{2})\alpha
L_{P}^{2}\,,$$ $$F=\Big(\frac{3}{2}+\frac{(D-1)}{2}\Big)\alpha
L_{P}^{2}.$$

Note that the first term in (21) gives the standard Wien's law in
the absence of quantum gravity effects. The extra terms in (21) are
quantum gravity corrections and as usual these corrections are
temperature dependent.
\section{Entropy and specific heat}
Now we calculate entropy and specific heat of black body radiation
in a model universe with $D$ spatial dimensions and in the presence
of quantum gravity effects encoded in MDR. We start with total
energy in cavity as
$$U(T)=V \rho(T)=V \frac{(D-1)2{\pi}^\frac{D}{2}
T^{D+1}}{\Gamma(\frac{D}{2})}\Bigg\{\zeta(D+1) \Gamma(D+1)$$
$$+\bigg(\frac{3}{2}+\frac{(D-1)}{2}\bigg)\alpha {L_p}^2
T^2\zeta(D+3)\Gamma(D+3)+$$
\begin{equation}
\bigg[\alpha^2\Big(-\frac{5}{8}+\frac{3}{2}\frac{(D-1)}{2}+\frac{(D-1)(D-3)}{8}+\alpha'\Big(\frac{5}{2}+\frac{D-1}{2}\Big)\bigg]{L_p}^4
T^4 \zeta(D+5)\Gamma(D+5)\Bigg\}\,.
\end{equation}
Then the specific heat and entropy of black body radiation are expressed respectively as\\
$$ C_V=\frac {\partial U }{\partial
T}\bigg{|}_{V=cte}=V\frac{(D-1)2{\pi}^\frac{D}{2}
}{\Gamma(\frac{D}{2})}\Bigg\{(D+1)T^{D} \zeta(D+1) \Gamma(D+1)$$
$$+\bigg(\frac{3}{2}+\frac{(D-1)}{2}\bigg)\alpha {L_p}^2 \zeta(D+3)
\Gamma(D+3)(D+3)T^{D+2}+$$
 \begin{equation}
\bigg[\alpha^2\Big(-\frac{5}{8}+\frac{3}{2}\frac{(D-1)}{2}+\frac{(D-1)(D-3)}{8}+\alpha'\Big(\frac{5}{2}+\frac{D-1}{2}\Big)\bigg]{L_p}^4
\zeta(D+5)\Gamma(D+5) (D+5) T^{D+4}\Bigg\},
\end{equation}
\\

$$S=\int_{0}^{T} \bigg(\frac{C_V}{T}\bigg) dT=V\frac{(D-1)2{\pi}^\frac{D}{2}
}{\Gamma(\frac{D}{2})}\Bigg\{\frac {(D+1)}{D}T^{D} \zeta(D+1)
\Gamma(D+1) +$$ $$\bigg(\frac{3}{2}+
 \frac{(D-1)}{2}\bigg)\alpha{L_p}^2 \zeta(D+3)
\Gamma(D+3)\frac{(D+3)}{(D+2)}T^{D+2}+$$
\begin{equation}
\bigg[\alpha^2\Big(-\frac{5}{8}+\frac{3}{2}\frac{(D-1)}{2}+\frac{(D-1)(D-3)}{8}+\alpha'\Big(\frac{5}{2}+\frac{D-1}{2}\Big)\bigg]{L_p}^4
\zeta(D+5)\Gamma(D+5) \frac {(D+5)}{(D+4)} T^{D+4}\Bigg\}\,.
\end{equation}

Figures $5$ and $6$ demonstrate the variation of specific heat
versus temperature without and with MDR respectively. As figure $5$
shows, the heat capacity decreases by increasing the number of
spatial dimensions. Comparison between figures $5$ and $6$ shows
that for a given temperature, the value of specific heat with
quantum gravity effects is larger than the corresponding value in
the absence of these effects. In other words, incorporation of
quantum gravity effects through MDR increases specific heat of black
body radiation for a given temperature.
\begin{figure}[htp]
\begin{center}
\includegraphics{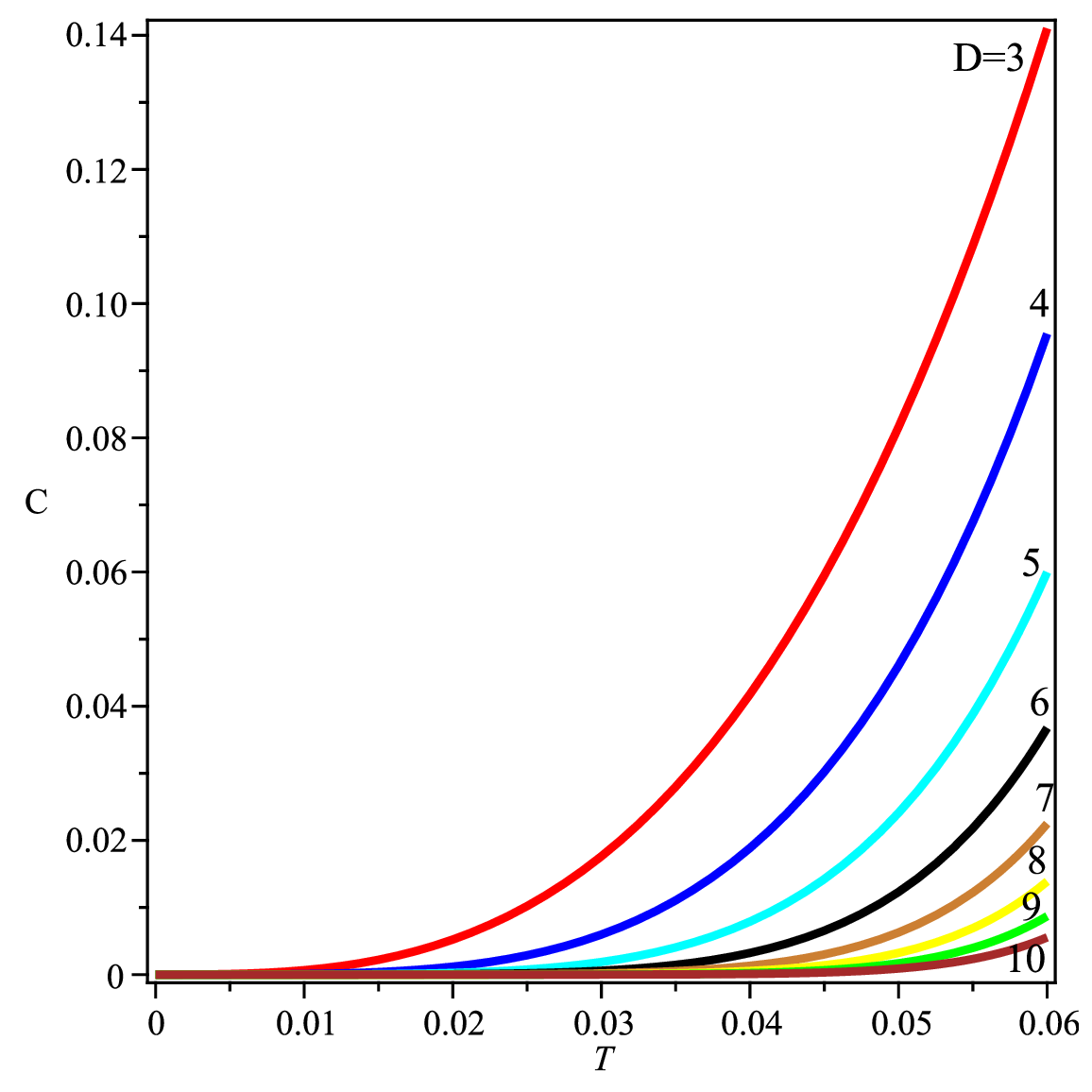}
\end{center}
\vspace{5.5cm} \caption{\scriptsize{Black body radiation specific
heat in a model universe with $D$ spatial dimensions without MDR
effects.}}
\end{figure}

\begin{figure}[htp]
\begin{center}
\includegraphics{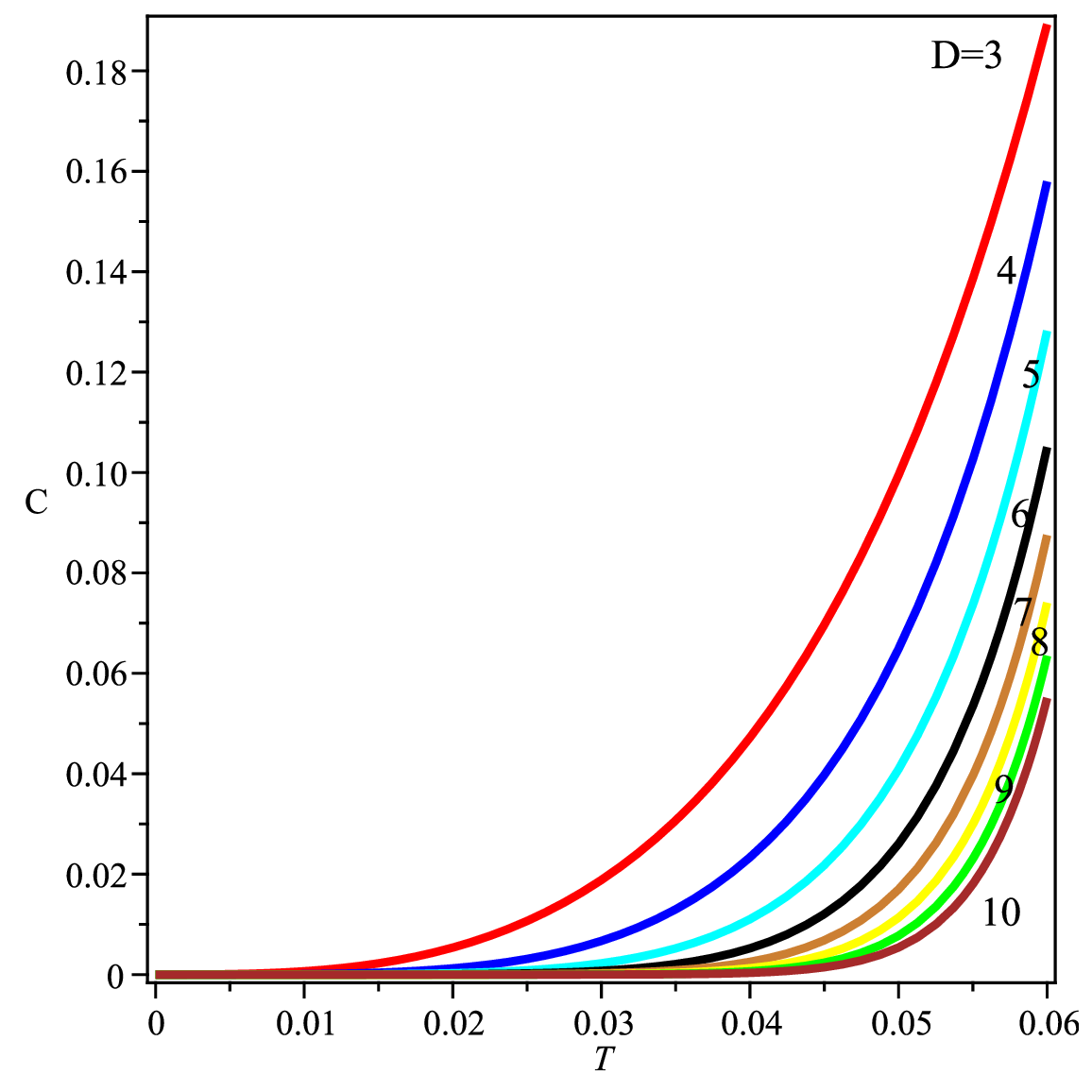}
\end{center}
\vspace{7cm} \caption{\scriptsize{Black body radiation specific heat
in a model universe with $D$ spatial dimensions in the presence of
MDR effects.}}
\end{figure}
\newpage
Figures $7$ and $8$ demonstrate the behavior of black body entropy
versus temperature for different numbers of spatial dimensions
without and with MDR effects respectively. Figure $7$ shows that the
entropy decreases as the number of spatial dimensions increases.
Comparison between figures $7$ and $8$ shows that quantum gravity
effects increase the entropy content of black body radiation in a
given temperature.

\begin{figure}[htp]
\begin{center}
\includegraphics{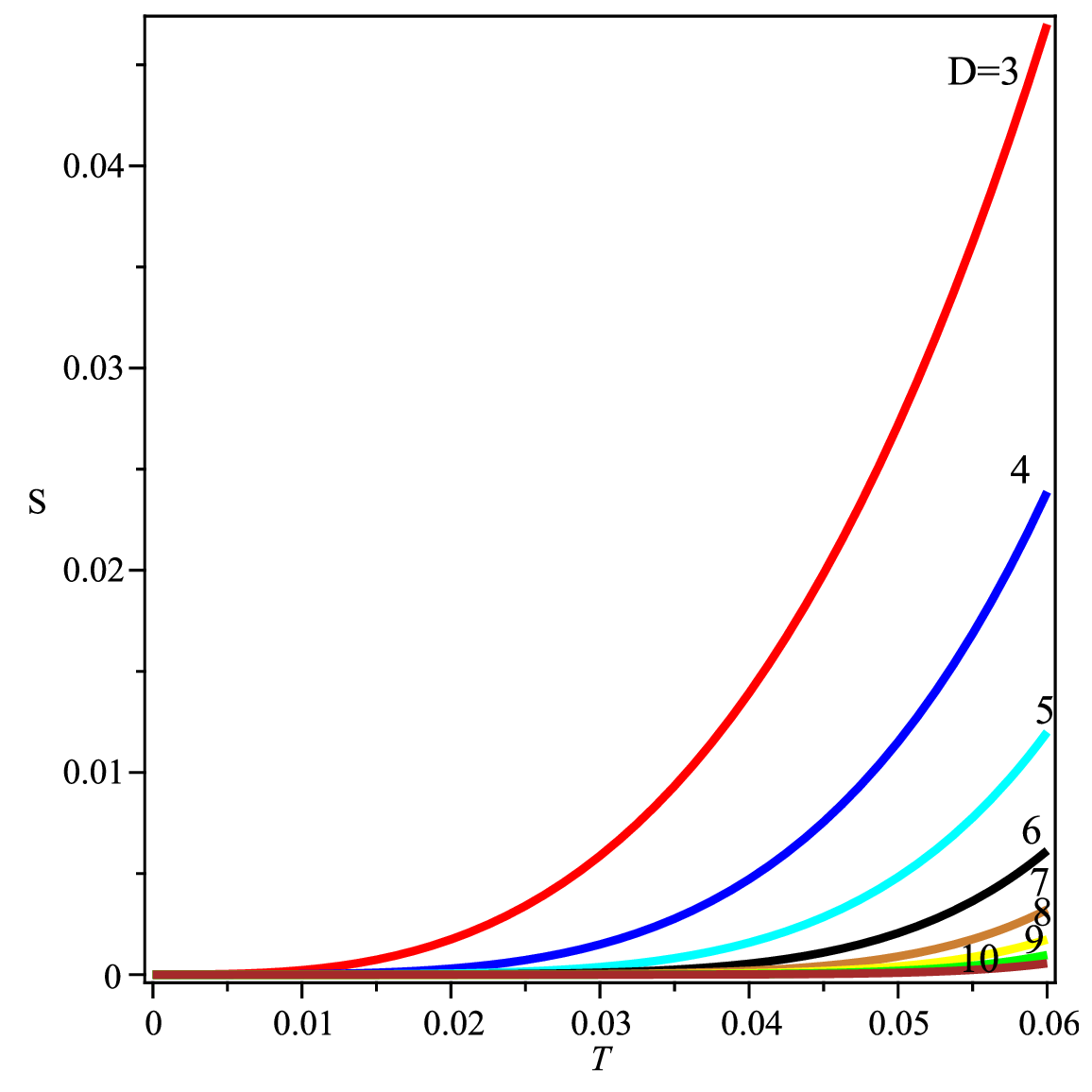}
\end{center}
\vspace{7cm} \caption{\scriptsize{Black body entropy in a model
universe with $D$ spatial dimensions without MDR effects.}}
\end{figure}

\begin{figure}[htp]
\begin{center}
\includegraphics{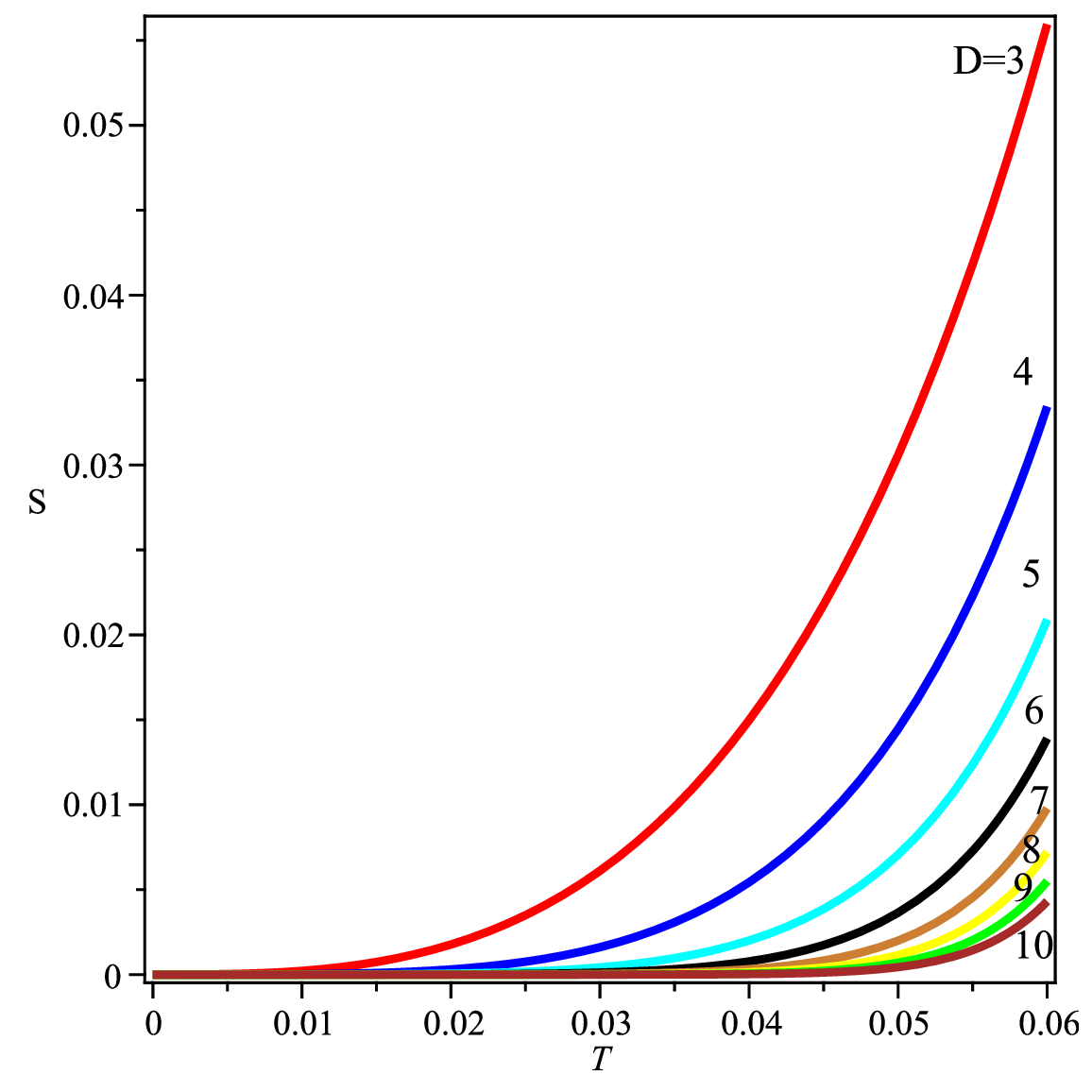}
\end{center}
\vspace{5cm} \caption{\scriptsize{Black body entropy in a model
universe with $D$ spatial dimensions in the presence of MDR
effects.}}
\end{figure}

\newpage

\section{Modified Debye's law}
Imagine a black body as a solid box, which is compound of vibrating
atoms. These vibrating atoms are recognized as the so-called phonons
in this box. The method adopted in this section resembles more or
less calculation of number of electromagnetic modes in black body
radiation. Nevertheless, there are some discrepancies between these
two cases since in a vibrating solid there are three types of waves
(one longitudinal and two transverse) in three spatial dimensions
while there are only two polarization states for electromagnetic
modes. So, if we generalize this situation to a model univerese with
$D$ spatial dimension, we would have $D$ possible polarization
states for phonons. Unlike photon, phonon frequency is bounded by
upper limit. In fact, the total vibrational states for phonons is
$D\times N$ where $N$ is the number of atoms. Since there is no
limitation on the number of phonons in each energy levels, phonon
could be considered as bosons. Therefore, in a model universe with
$D$ spatial dimensions, we have
\begin{equation}
\rho(\nu) d\nu=D L^{D} \frac{2\pi^{\frac{D}{2}}}
{\Gamma(\frac{D}{2})} \int_{0}^{\nu_{max}} \frac
{\nu^{D}}{(e^\frac{\nu}{T}-1)} d\nu,
\end{equation}
which determines the energy density of vibrating solid in a model
universe with $D$ spatial dimensions. By setting
$x=\frac{\nu}{T}$, we obtain \\
\begin{equation}
\rho(\nu) d\nu=D^{2} T N \bigg(\frac{T}{T_D}\bigg)^{D}
\int_{0}^{x_{max}} \frac{x^{D}}{e^{x}-1}dx,
\end{equation}
where $N$ is the number of atoms in vibrating solid and $T_D$ is
introduced as Debye temperature which is equal to
$$\Big(T_{D}\Big)^{D}=\frac {N\Gamma(\frac{D}{2}) D} {2 V
\pi^{\frac{D}{2}}}.$$

Now, we incorporate quantum gravity effects through MDR. Based on our
previous analysis, we have
 $$ \rho(x) dx=D^{2} T N \bigg(\frac{T}{T_D}\bigg)^{D} \int_{0}^{x_{max}}
\frac{x^{D}}{e^{x}-1}\Bigg\{1+\bigg(\frac{3}{2}+\frac{(D-1)}{2}\bigg)\alpha
L_p^2 (xT)^2+$$
\begin{equation}
\bigg[\alpha^2\Big(\frac{-5}{8}+\frac{3}{2}\frac{(D-1)}{2}+\frac{(D-1)(D-3)}{8}\Big)
+\alpha'\Big(\frac{5}{2}+\frac{(D-1)}{2}\Big)\bigg]L_{p}^4 (xT)^4\Bigg\}dx.
\end{equation}
The temperature of a Debye solid is said to be low if $T \ll T_D$.
In this limit we find \\
$$U(T)=D^2 T N\bigg(\frac{T}{T_D}\bigg)^{D} \Bigg\{\zeta(D+1) \Gamma(D+1)+\bigg(\frac{3}{2}+\frac{(D-1)}{2}\bigg)\alpha {L_p}^2
T^2 \zeta(D+3) \Gamma(D+3)$$
\begin{equation}
+\bigg[\alpha^2\Big(-\frac{5}{8}+\frac{3}{2}\frac{(D-1)}{2}+\frac{(D-1)(D-3)}{8}\Big)
+\alpha'\Big(\frac{5}{2}+\frac{D-1}{2}\Big)\bigg]{L_p}^4 T^4
\zeta(D+5)\Gamma(D+5)\Bigg\}\,.
\end{equation}
By differentiating this relation with respect to $T$, we find \\
$$C_V=D^2(D+1)  N\bigg(\frac{T}{T_D}\bigg)^{D} \zeta(D+1) \Gamma(D+1)$$ $$+\bigg(\frac{3}{2}+\frac{(D-1)}{2}\bigg)\alpha
{L_p}^2 D^2 N\frac{T^{D+2}}{(T_D)^{D}}(D+3) \zeta(D+3) \Gamma(D+3)$$
\begin{equation}
+\bigg[\alpha^2\Big(-\frac{5}{8}+\frac{3}{2}\frac{(D-1)}{2}+\frac{(D-1)(D-3)}{8}\Big)
+\alpha'\Big(\frac{5}{2}+\frac{D-1}{2}\Big)\bigg]{L_p}^4 D^2 N (D+5)
\frac{T^{D+4}}{(T_D)^{D}} \zeta(D+5)\Gamma(D+5).
\end{equation}
This is the modified Debye law in the MDR framework. Note that the
first term in the right hand side gives the standard Debye law in
the absence of quantum gravity effects. It's worth mentioning to
note that modified specific heat changes with temperature. Also
quantum gravity corrections are temperature dependent as usual.

\section{Modified Dulong-petit law}

Dulong and Petit found that the heat capacity of a mole of many
solid substances is about $3R$, where $R$ is the universal gas
constant. An equivalent statement of the Dulong-Petit law in modern
terms is that, regardless of the nature of the substance or crystal,
the specific heat capacity $C$ of a solid substance (measured in
Joule per Kelvin per Kilogram) is equal to $\frac{3R}{M}$ where $M$
is the molar mass (measured in Kilogram per Mole). Thus, the heat
capacity per mole of many solids is $3R$. Actually this statement is
valid only at high temperature and in this case the specific heat is
assumed to be temperature-independent in classical viewpoint. Now
we reconsider this model in the MDR framework in order to see how
quantum gravity effects modify the classical Dulong-Petit law. At
high temperature when $T\gg T_{D}$, we can use the approximation
$e^{x}-1\approx x$. Therefore, based on Eq. (27), the total energy
of vibrating solid is given by

 $$U(T)=D T N+\frac {D^{2}}{(D+1)} N
T_{D}\bigg(\frac{3}{2}+\frac{(D-1)}{2}\bigg)\alpha L_p^2
T^2+\bigg(\alpha^2(\frac{-5}{8}+\frac{3}{2}\frac{(D-1)}{2}+\frac{(D-1)(D-3)}{8})$$
\begin{equation}
+\alpha'(\frac{5}{2}+\frac{(D-1)}{2})\bigg)L_{p}^4
\frac{D^{2}}{(D+3)^{2}} N  T^{2} T_D^{3}\,.
\end{equation}
By differentiating with respect to $T$, we find
 $$C_{V}=D  N+\frac {2D^{2}}{(D+1)} N
T_{D}\bigg(\frac{3}{2}+\frac{(D-1)}{2}\bigg)\alpha L_p^2
T+2\bigg[\alpha^2\Big(\frac{-5}{8}+\frac{3}{2}\frac{(D-1)}{2}+\frac{(D-1)(D-3)}{8}\Big)$$
\begin{equation}
+\alpha'\Big(\frac{5}{2}+\frac{(D-1)}{2}\Big)\bigg] L_{p}^4
\frac{D^{2}}{(D+3)^{2}} N  T T_D^{3}.
\end{equation}
This is the modified Dulong-petit law. In the absence of quantum
gravity effects, ordinary quantum mechanics result will be achieved,
that states that the heat capacity is independent of temperature.
However, in the quantum gravity domain one cannot ignore temperature
dependence of heat capacity as is evident in correction terms. These
temperature-dependent quantum gravity corrections reveal some
phenomenological aspects of quantum gravity proposal. These features
may provide also a direct basis to test quantum gravity in the lab.

\newpage
\section{Conclusion}
Taking into account the MDR corrections in a model universe with large
extra dimensions would be helpful in providing an outlook in the
study of black body radiation. How can MDR affect black body radiation in a model universe with extra dimensions? In studying the black body Planck distribution in the absence of MDR effects, it has been shown that the frequency at which the distribution maximizes moves toward higher values by increasing the number of spatial dimensions. Moreover, Planck distribution becomes wider and its height decreases by increasing $D$. In the presence of MDR effects, the location of peaks moves to higher frequencies by increasing $D$ while the height and the width of peaks increases. It seems that the MDR effects provide a significant corrections into Planck distribution. The point that should be stressed here is that the quantum gravitational effects are enhanced as the number of spatial dimensions increases.
Rayleigh-Jeans law, Stefan-Boltzmann law and Wien's law are modified within MDR. The presence of temperature in the modified Stefan-Boltzmann law is worth
reiterating at this point since the temperature-dependent of
correction terms stem from the fact that a smooth continues space
breaks down in quantum gravity regime and it produces some
fluctuations that cause the turbulence of space-time. Another
interesting point is related to the jeans's number which increases
in the presence of quantum gravity effects. Moreover, the values of entropy and specific heat of black body
radiation within the consideration
of MDR are larger than the ones in the absence of MDR. In studying the modified Debye's law within MDR, it has been concluded that the modified specific heat changes with temperature. In addition, if one tries to investigate the Dulong-Petit law within MDR in a model
universe with large extra dimensions, the direct effects of
temperature as an important character of correction terms are
noticeable. The appearance of
temperature-dependent correction terms may be related to some
phenomenological aspects of quantum gravity. Of course, one can utilize these modified features to test the quantum gravity in the lab.\\

{\bf Acknowledgment}\\

The work of K. Nozari has been supported financially by Research
Institute for Astronomy and Astrophysics of Maragha (RIAAM) under
research project number 1/4717-**.

\end{document}